\documentclass[12pt]{spieman}  % 12pt font required by SPIE;
\usepackage{amsmath,amsfonts,amssymb}
\usepackage{graphicx}
\usepackage{setspace}
\usepackage{tocloft}

\title{Autofluorescence spectroscopy for determining cell confluency}

\author[a,*]{Derrick Yong}
\author[b]{Ahmad Amirul Abdul Rahim}
\author[b]{Jesslyn Ong}
\author[b,${\dagger}$]{May Win Naing}
\affil[a]{Precision Measurements Group, Singapore Institute of Manufacturing Technology, 2 Fusionopolis Way, Innovis \#08-04, Singapore, 138634}
\affil[b]{Bio-Manufacturing Programme, Singapore Institute of Manufacturing Technology, 2 Fusionopolis Way, Innovis \#08-04, Singapore, 138634}

\cftpagenumbersoff{figure}
\cftpagenumbersoff{table} 
\begin{document} 
\maketitle

\begin{abstract}
%no more than 200 words
Patient-specific therapies require that cells be manufactured in multiple batches of small volumes, making it a challenge for conventional modes of quality control. The added complexity of inherent variability (even within batches) necessitates constant monitoring to ensure comparable end products. Hence, it is critical that new non-destructive modalities of cell monitoring be developed. Here, we study, for the first time, the use of optical spectroscopy in the determination of cell confluency. We exploit the autofluorescence properties of molecules found natively within cells. By applying spectral decomposition on the acquired autofluorescence spectra, we are able to further discern the relative contributions of the different molecules, namely flavin adenine dinucleotide (FAD) and reduced nicotinamide adenine dinucleotide (NADH). This is then quantifiable as redox ratios that represent the extent of oxidation to reduction based upon the optically measured quantities of FAD and NADH. Results show that RR is significantly higher for lower confluencies ($\leq$50$\%$), which we attribute to the different metabolic requirements as cells switch from individual survival to concerted proliferation. We validate this relationship through bio-chemical assays and autofluorescence imaging, and further confirmed through a live-dead study that our measurement process had negligible effects on cell viability.
\end{abstract}

% Include a list of up to six keywords after the abstract

% Include email contact information for corresponding author
{\noindent \footnotesize\textbf{*}Derrick Yong,  \linkable{derrick-yong@SIMTech.a-star.edu.sg}; {$^{\dagger}$}May Win Naing,  \linkable{winnaingm@SIMTech.a-star.edu.sg}; }

\begin{spacing}{2}   % use double spacing for rest of manuscript

\section{Introduction}
\label{sect:intro}  

The cell therapy industry has garnered significant momentum in recent years, pivoting on the promise that cell-based therapies hold in treating conditions where conventional approaches have failed. As therapies make the leap from lab to bedside, a major challenge highlighted in the manufacturing of such therapies lies with establishing quality and developing control processes\cite{Lipsitz2016,Roh2016}. With patient-specific therapies, there is added complexity as a result of the inherent variability of cells (donor-to-donor variation). Destructive testing of such therapies is time-consuming, costly and essentially reduces the available dosage for the patient. Ideally, monitoring methods to ensure quality of such products should be achievable \textit{in situ}, non-destructively and in real-time. Optical spectroscopy offers a solution to said requirements\cite{Teixeira2009,ClaBen2017}, with effects like Raman scattering\cite{Butler2016} and autofluorescence\cite{Croce2017} offering specificity under a label-free modality.

Cells contain bio-molecules capable of emitting fluorescence, which is known as cell autofluorescence\cite{NaturalBiomarkers2014}. These cell-endogenous fluorophores are the very same bio-molecules responsible for the host of cellular processes that govern cell functions and metabolic activities. Different fluorophores can be differentiated by their spectral distribution of emissions, with the amount of emission further corresponding to their respective quantities. Since the pioneering work by Chance \textit{et al.}\cite{Chance1962}, where a relationship was established between cell autofluorescence and cellular metabolic processes, cell-endogenous fluorophores have been successfully used as biomarkers in the non-destructive and real-time determination of cell characteristics. Numerous adoptions have thus been made in biomedical research and diagnosis\cite{Croce2014}, with notable applications in the identification of stem cell differentiation\cite{Rice2010,Quinn2013} as well as the detection of diseases such as cancer\cite{Skala2007,Miller2017} and Alzheimer's\cite{Shi2017}. These applications have been enabled by optical techniques such as multi-photon microscopy cum spectroscopy\cite{Huang2002,Quinn2013} and fluorescence lifetime imaging microscopy\cite{Lakowicz1992,Blacker2014}. Aside from these capital- and skill-intensive techniques, which offer in depth details more pertinent to the fundamental understanding of the biosciences, more broadly adoptable and economical methods like multispectral microscopy\cite{Gosnell2016} and autofluorescence spectroscopy \cite{Croce1999,Croce2017} have also been reported as practical alternatives. 

One measurand of interest is the metabolic state of cells\cite{Huang2002,Heikal2010,Georgakoudi2012,Croce2017}, which offers a direct indication of the cells' activity and could hence serve as a means of monitoring cells during the manufacturing process. This is quantified using a redox ratio (RR) that indicates the extent of oxidation against reduction based upon the optically measured amounts of metabolic co-enzymes --- flavin adenine dinucleotide (FAD) and reduced nicotinamide adenine dinucleotide (NADH), correspondingly. In the context of cell manufacturing, knowing when to passage (or sub-culture) cells is crucial to optimizing output and establishing quality control. This decision is conventionally dependent upon a human operator's judgement of cell confluency, which refers to the area coverage of cells in a culture vessel, but has more recently progressed towards automation through studies involving vision-based systems\cite{Ker2011}. Such systems are however limited by a millimeter field of view, and although automation and scanning mechanisms are capable of extending it, the complexity and higher costs poses challenges to scalability in cell manufacturing. Spectroscopy, on the other hand, can be performed using evanescent excitations that can readily be scaled up using optical waveguides \cite{Grandin2006}, where the excitation area can be arbitrarily wide and limited only by the optical waveguide design. In this work, we studied the use of autofluorescence spectroscopy as an alternative method for the determination of cell confluency. We acquired autofluorescence spectra from cells at different confluencies and spectrally decomposed them to determine RR. A relationship was then established between cell confluency and RR.

\section{Methods and Materials}
\label{sect:methods}  

\begin{figure}
\begin{center}
\begin{tabular}{c}
\includegraphics[height=12cm]{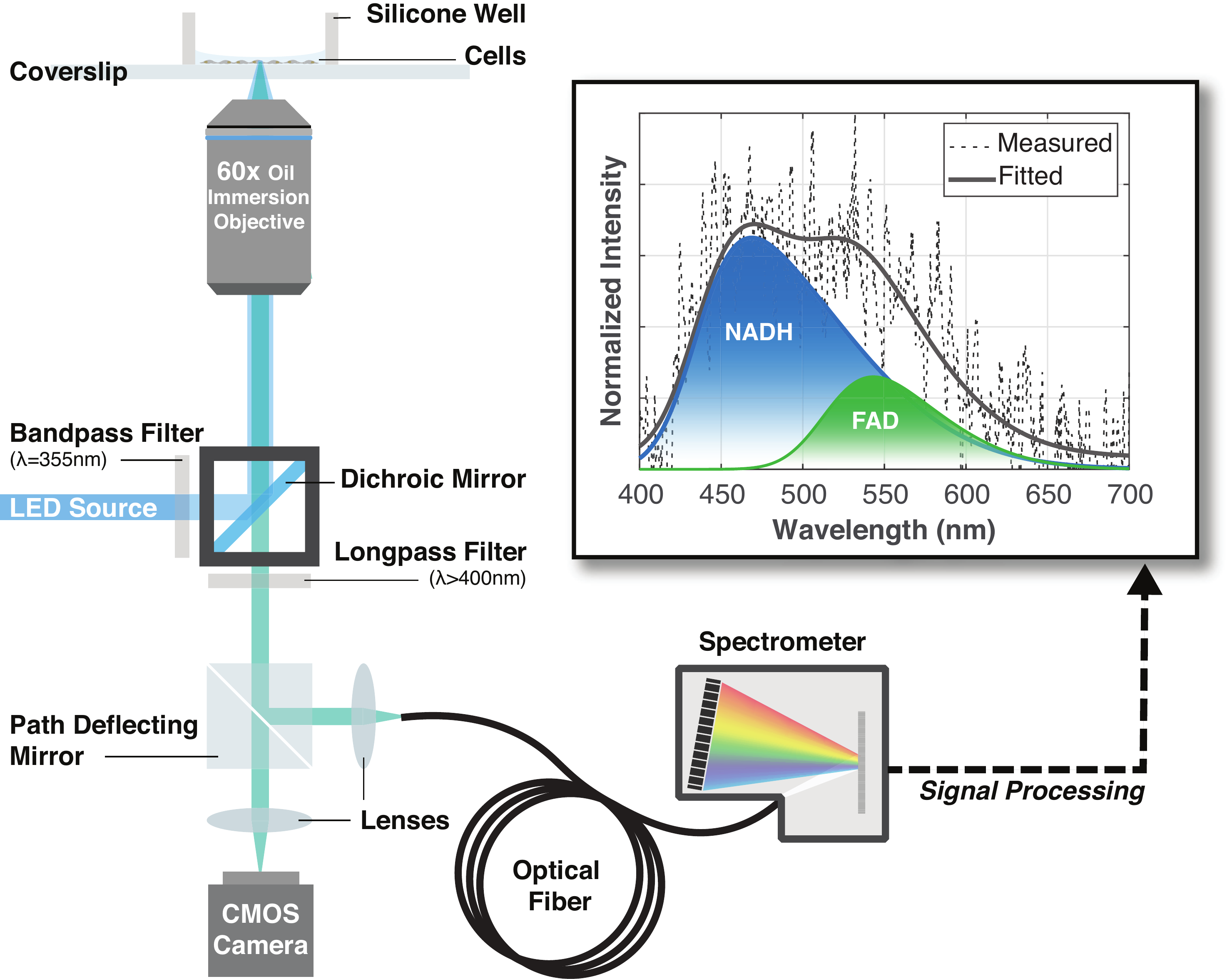}
\end{tabular}
\end{center}
\caption 
{ \label{fig:Schematics}
Schematics of microspectroscopy setup. Setup is based upon an inverted fluorescence microscope with the addition of a custom fluorescence filter cube and an attached spectrometer. Cells adhered to glass coverslips are immersed in PBS during measurements. Inset: Autofluorescence spectra collected from live cells. The measured and fitted spectra is represented by a black-dotted and a solid-grey curve, correspondingly. The constituent autofluorescence emissions are indicated by blue and green curves for the emissions of NADH and FAD respectively.} 
\end{figure}

\subsection{Cell culture}

WS1 fibroblast cells (CRL2029, ATCC, USA) at Passage 5 were thawed. Each vial, consisting of approximately 1$\times$10$^{6}$cells, was added to 9ml of culture media formulated using 10$\%$ Fetal Bovine Serum (HyClone$\texttrademark$, GE Healthcare Life Sciences, USA) and 90$\%$ Minimum Essential Medium Eagle with Earle's Balanced Salt Solution and L-Glutamine (PAN-Biotech, Germany). Cells were centrifuged at 130$\times$g at 4$^{\circ}$C for 5mins, and the resulting supernatant was discarded. Cells were then re-suspended in 10ml fresh culture media by aspirating with a 10ml serological pipette (CELLSTAR$^{\textregistered}$, Gernier, Germany). The cell suspension was then transferred to a 75cm$^{2}$ cell culture flask (Corning$^{\textregistered}$, USA). The culture flask was next placed into a CO$_{2}$ incubator (Forma Steri-Cycle i160, Thermo Fisher Scientific, USA) set at 37$^{\circ}$C, 95$\%$ humidity and 5$\%$ CO$_{2}$; and left to incubate for 3 to 4 days before being harvested for seeding. 

Cells in the culture flask were washed twice with 5ml of Phosphate Buffered Saline (HyClone$\texttrademark$, GE Healthcare Life Sciences, USA) before 2ml of 0.25$\%$ Trypsin:EDTA solution (Gemini Bio-Products, USA) was introduced. After 4min, 20ml of culture media was added and all the contents of the culture flask were extracted for centrifugation at 130$\times$g and 4$^{\circ}$C for 5mins. The resulting supernatant was discarded and the cells were re-suspended in 5ml of culture media. Cell concentration was determined with an automated cell counter (EVE$\texttrademark$ Automated Cell Counter, NanoEnTek, South Korea). 20$\mu$l of cell suspension was mixed with an equal volume of Trypan Blue and transferred into cell counting slides (EVE$\texttrademark$ Chamber Slides, NanoEnTek, South Korea) before being inserted into the automated cell counter. Cell suspension were then diluted to the required concentrations prior to seeding for microspectroscopy and the bio-chemical assays.

All cell culture work and related procedures were performed within a Biosafety Cabinet (1300 Series A2, Thermo Fisher Scientific, USA).

\subsection{Autofluorescence microspectroscopy}

Microspectroscopy was performed using an inverted fluorescence microscope (IX73, Olympus, Japan) after a simple upgrade. Schematics are shown in Fig. \ref{fig:Schematics}. Excitation was supplied by a LED illumination source (wLS, QImaging, Canada) with a center wavelength of 365nm. In order to excite and collect cell autofluorescence, a customized fluorescence filter cube was assembled with optics obtained from Thorlabs (USA). This comprised an excitation bandpass filter with a center wavelength of 355nm and FWHM of 10nm (FLH355-10); an emission longpass filter with a cut-on wavelength of 400nm (FELH0400); and a dichroic mirror that reflects wavelengths below 407nm and transmits wavelengths above 425nm (MD416). Spectral measurements were obtained with a USB-linked spectrometer (Maya2000, Ocean Optics, USA) that was attached to one of the microscope's camera port via a collimator (CVH100-COL, Thorlabs, USA) and a 600$\mu$m core optical fiber patch cable (QP600-2-VIS-NIR, Ocean Optics, USA). 

For autofluorescence microspectroscopy, cells were seeded on 0.175mm-thick glass coverslips (D263M, Schott, Germany) within silicone wells with growth areas of $\sim$2cm$^{2}$ and heights of $>$10mm. Silicone wells were fabricated from medical grade silicone (Silpuran 6000/10, Wacker Chemie AG, Germany). 1ml of culture media was used to wash the silicone wells prior to the seeding of cells. Cells were seeded at a concentration of 1$\times$10$^{4}$cells/ml and 1ml was transferred to each coverslip within the confines of their silicone wells. The coverslips were then incubated within the CO$_{2}$ incubator and left undisturbed until extracted for autofluorescence microspectroscopy. Each coverslip was taken out of the incubator for microspectroscopy at 24hr-intervals over the total duration of up to 7 days or when cells were at 100$\%$ confluency. Prior to microspectroscopy, culture media in the silicone well was removed and the cells were washed with 1ml of Phosphate Buffered Saline (PBS) with Ca$^{2+}$ and Mg$^{2+}$ (HyClone$\texttrademark$, GE Healthcare Life Sciences, USA). 100$\mu$l of the same buffer was then added. Coverslips were then mounted on the stage of the fluorescence microscope for microspectroscopy.

Autofluorescence spectra were obtained by first locating the region of interest on the coverslip through the microscope's camera before deflecting the collected light to the spectrometer. Each spectral measurement was recorded over the integration time of 10s. This was done through a 60$\times$ oil-immersion super apochromat objective (UPLSAPO60XO, Olympus, Japan) together with a low autofluorescence immersion oil (IMMOIL-F30CC, Olympus, Japan). To minimize the exposure of cells to the excitation source, we first selected the region of interest under low bright field illumination before switching to the LED illumination for image capture and spectral measurements. For each sample, at least ten measurements (N$\geq$10) were made at different locations on the cell-covered regions of the coverslip. This was followed by background measurements of a 100$\mu$l drop of PBS with Ca$^{2+}$ and Mg$^{2+}$ at a clean region of the coverslip. Background measurements were repeated five times per coverslip. We repeated the measurements over six sets of samples to obtain at least three sets of data per reported confluency.

\subsection{Spectral decomposition and optical redox ratio calculation}

Collected autofluorescence spectra were processed using a MATLAB-based software developed in our laboratory. The software is designed to perform signal processing, background correction and spectral decomposition that breaks down a spectrum into its constituent emission spectra. The latter was based on a non-linear curve-fitting procedure by Croce \textit{et al.} \cite{Croce1999,Croce2017}. In this work, we limited the spectral decomposition to just two cell-endogenous fluorophores of interest - reduced nicotinamide adenine dinucleotide (NADH) and flavin adenine dinucleotide (FAD). Spectral fitting parameters for NADH and FAD were obtained by fitting emission spectra of corresponding reference solutions acquired through the same fluorescence microscope configuration. A plot illustrating the result of said spectral decomposition is shown in Fig. \ref{fig:Schematics}(inset). 

From each autofluorescence spectrum, a redox ratio (RR) was computed as [FAD]/([FAD] + [NADH]). The concentration of each cell-endogenous fluorophore (Fl) is related to its total fluorescence emission by:
\begin{equation} \label{eq:fluoconc}
	[Fl] = \frac{\int I_{Fl} d\lambda}{I_0 \epsilon_{Fl} \phi_{Fl} L}
\end{equation}
\noindent where $\int I_{Fl} d\lambda$ is the sum of intensities over the entire wavelength span of the fluorophore's emission; $I_0$ is the input excitation intensity; $\epsilon_{Fl}$ is the fluorophore's extinction coefficient at the excitation wavelength; $\phi_{Fl}$ is the fluorophore's quantum yield; and L is the path length of interaction between the input excitation and fluorophore. Substituting this into the RR and simplifying gives an optical variation of the RR:
\begin{equation} \label{eq:redoxratio}
	RR = \frac{\int I_{FAD} d\lambda\times\epsilon_{NADH} \phi_{NADH} }{\int I_{FAD} d\lambda\times\epsilon_{NADH} \phi_{NADH} + \int I_{NADH} d\lambda\times\epsilon_{FAD} \phi_{FAD} }
\end{equation}
\noindent where $\epsilon_{NADH}$ and $\epsilon_{FAD}$ were experimentally determined to be $\sim$5000M$^{-1}$cm$^{-1}$ and $\sim$8000M$^{-1}$cm$^{-1}$ respectively, at an excitation wavelength of 355nm; $\phi_{NADH}$ and $\phi_{FAD}$ were 0.019\cite{Scott1970} and 0.033\cite{Islam2003}.

\subsection{Live-dead study}

A cell viability assay kit (LIVE/DEAD$\texttrademark$ Viability/Cytotoxicity Kit, Invitrogen, USA) was used in the determination of live and dead cells. Stained cells were imaged using the same fluorescence microscope under 4$\times$ magnification. Live and dead cells were identified through the standard FITC and TRITC filter cubes, correspondingly. The images were processed using ImageJ and cell viability was calculated as (number live cells)/(number of live cells + number of dead cells)$\times$100$\%$. The live-dead study was conducted on samples following microspectroscopy. This was done together with a control that was subjected to the exact same treatments less any form of illumination in the fluorescence microscope.

\subsection{Bio-chemical assays}

Two bio-chemical assays were conducted for the quantification of NADH and FAD using the NAD/NADH-Glo\texttrademark Assay Kit (Promega Corporation, USA) and FAD Assay Kit (Abcam, United Kingdom), correspondingly. For each assay, cells were seeded at a concentration of 3$\times$10$^{4}$cells/ml, where 30ml of cell suspension was transferred to each of the seven 175cm$^{2}$ cell culture flask (Corning$^{\textregistered}$, USA). The culture flasks were then incubated within the CO$_{2}$ incubator and left undisturbed until extracted for the assay. 

For each flask, multiple measurements (N$\geq$4) were performed on the same cell sample at different dilutions. Each cell sample was measured concurrently with a set of standards so as to ascertain the total intracellular concentrations for both NADH and FAD. The intracellular concentrations per cell were subsequently determined for each measurement by accounting for the different extents of dilutions.

\section{Results and Discussion}
\label{sect:results}  

\begin{figure}
\begin{center}
\begin{tabular}{c}
\includegraphics[width=\linewidth]{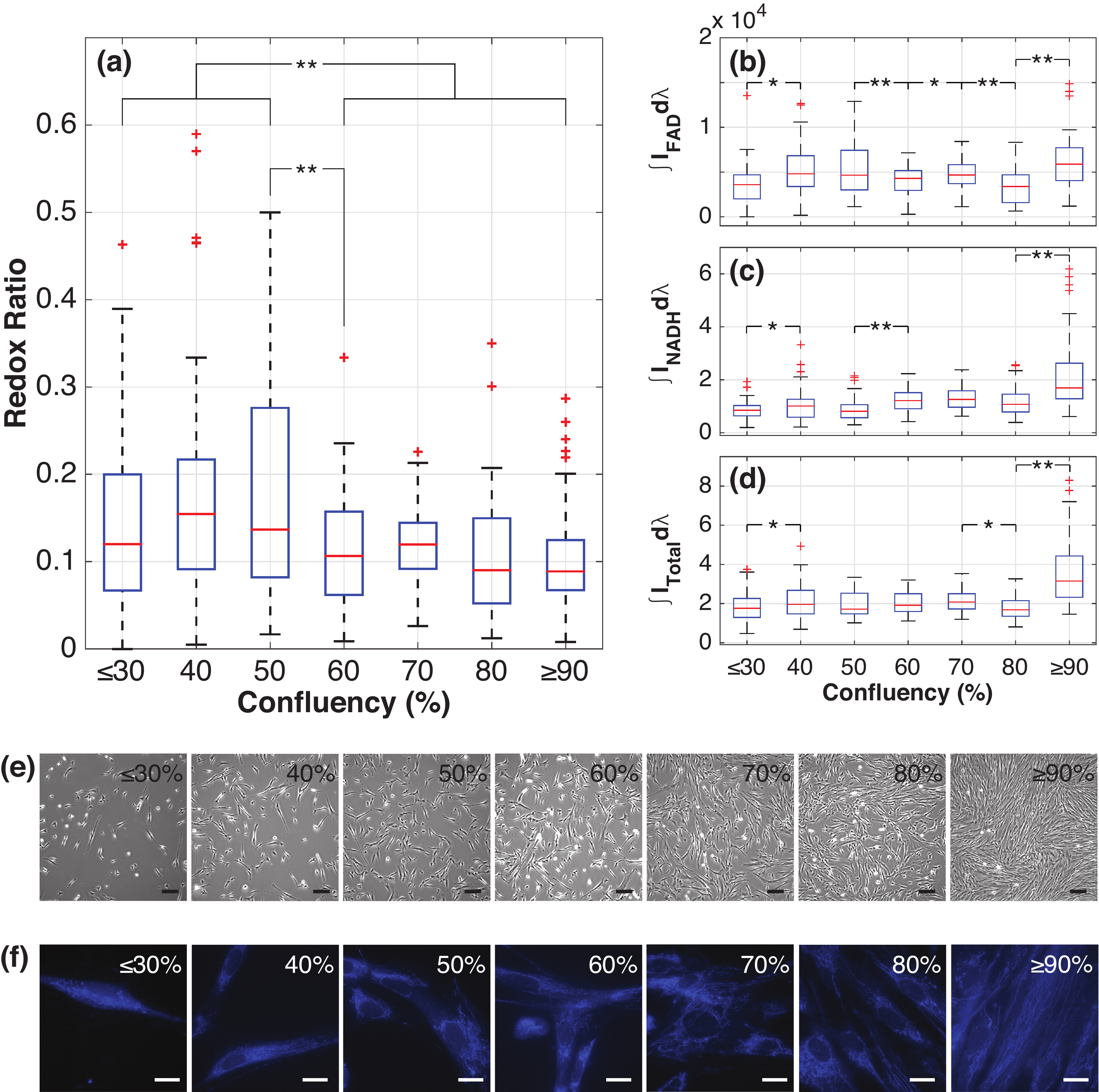}
\end{tabular}
\end{center}
\caption 
{ \label{fig:ResultsCombined}
Box plot representations of (a) redox ratios and (b--d) sum of intensities for varying confluencies of fibroblasts cultured on glass coverslips. Corresponding (e) phase contrast and (f) autofluorescence images of fibroblasts at different confluencies. Intensities over the entire wavelength span were summed for the constituent emissions of (b) FAD and (c) NADH as well as (d) the total autofluorescence emission. Each confluency interval comprises at least ten different spectral measurements acquired from at least three of the six different sets of samples. Red horizontal lines within the box plots correspond to the median values. Red `+'s represent data points are outliers. Statistical analysis: Data pairs marked with `*' and `**' are statistically significant based on a two-tailed Student's t-test --- P$<$0.05 and P$<$0.001 respectively. Scale bars: Black and white bars represent 200$\mu$m and 20$\mu$m, correspondingly.} 
\end{figure} 

We calculated redox ratios (RR) for each of the acquired autofluorescence spectra and organized them in Fig. \ref{fig:ResultsCombined}a. The sum of intensities for FAD ($\int I_{FAD} d\lambda$) and NADH ($\int I_{NADH} d\lambda$) emissions, used in the calculations of RR in Eq. \ref{eq:redoxratio}, were likewise collated in Fig. \ref{fig:ResultsCombined}b and c respectively. This was similarly done for the sum of intensities for the total autofluorescence ($\int I_{Total} d\lambda$) emission in Fig. \ref{fig:ResultsCombined}d. Data were grouped according to the respective confluencies of the cell samples they were collected from. Confluencies were determined through a side-by-side comparison of phase contrast images captured at 4x magnification as shown in Fig. \ref{fig:ResultsCombined}e. This was done at an accuracy of 10$\%$ between 40$\%$ and 80$\%$. It was however difficult to discern between the very low confluencies of $\leq$30$\%$ and the near-maximum confluencies of $\geq$90$\%$. Hence, data within these spans were further clustered. 

Each autofluorescence spectrum was signal processed and decomposed into the emission spectra of two cell-endogenous fluorophores --- NADH and FAD. Spectral decomposition was based upon the method detailed by Croce \textit{et al.} \cite{Croce1999,Croce2017}. In brief, we fitted the collected spectrum with two asymmetric Gaussian curves with peaks at approximately 470nm and 540nm, corresponding to the emissions of NADH and FAD measured in the same optical configuration. A typical decomposition is shown in Fig. \ref{fig:Schematics}(inset). We note that in the work by Croce \textit{et al.} multiple emission spectra ($\geq$4) were used to achieve fits with R-squared (R$^2$) values of $\sim$0.99. In this work we simplified the fitting to just the two key cell-endogenous fluorophores, and made the assumption of negligible spectral differences between free and bound NADH. In doing so, we noted average R$^2$ values of $>$0.75. Although less perfect fits were obtained, we find it sufficient in computing simple ratiometric relationships such as the redox ratio. This is especially true if we have prior knowledge on the primary emissions contributing to the spectrum. 

\begin{figure}
\begin{center}
\begin{tabular}{c}
\includegraphics[height=6cm]{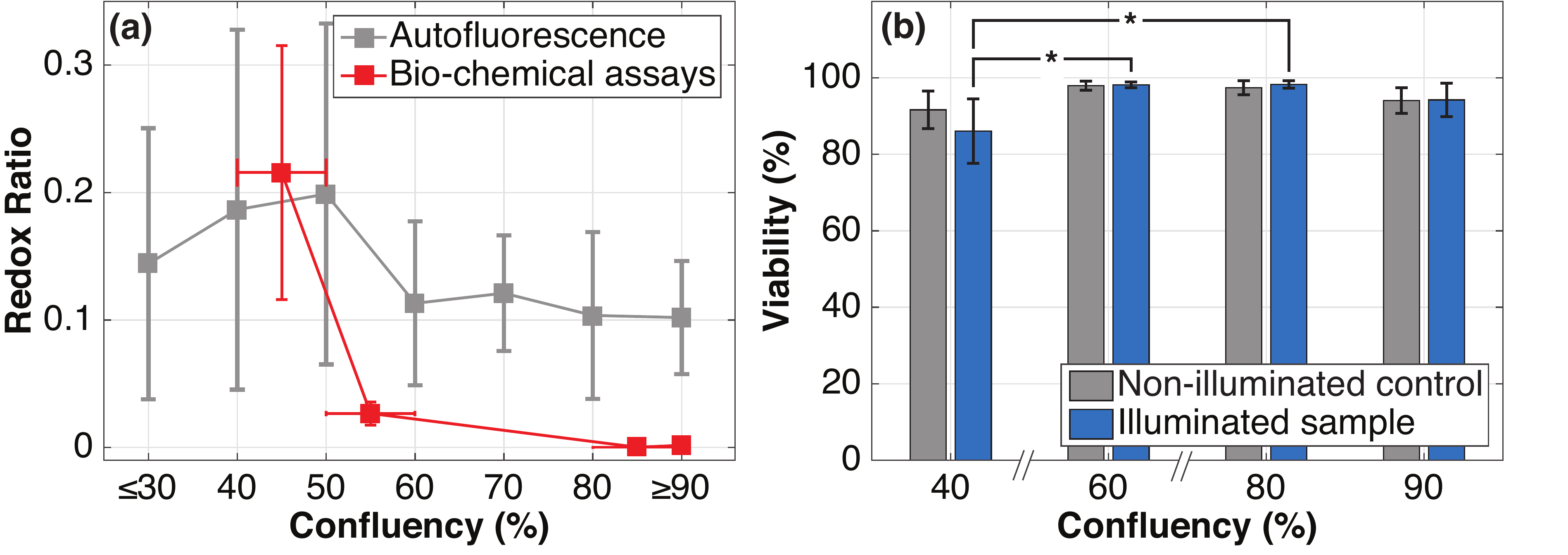}
\end{tabular}
\end{center}
\caption 
{ \label{fig:Assays}
Assay results for (a) redox ratios and (b) viability studies. (a) Comparison between redox ratios determined from bio-chemical assays (in red) and autofluorescence microspectroscopy (in grey) for varying confluencies of fibroblasts cultured on glass coverslips. (b) Comparison between cell viability for fibroblasts following microspectroscopy (Illuminated sample) and corresponding fibroblasts that were treated the same but without any form of illumination within the fluorescence microscope (Non-illuminated control). Presented data was obtained from the same batch of culture over 4 days. Statistical analysis: Data pairs marked with `*' are statistically significant based on a two-tailed Student's t-test --- P$<$0.05}
\end{figure}

Data are presented with boxplots in Fig. \ref{fig:ResultsCombined} to highlight their spread, which also reflect the inherent variabilities of cell samples (the same data are also presented in a mean and standard deviation plot in Fig. \ref{fig:Assays}). In Fig. \ref{fig:ResultsCombined}a, we note that RR extends up to 0.5 for confluencies of 50$\%$ and below, while hardly reaching 0.3 for higher confluencies. Statistical analysis comparing data from confluencies of $\leq$50$\%$ against data from confluencies $>$50$\%$ show a statistical significance --- P$<$0.001 using a two-tailed Student's t-test. Further statistical analysis also shows that data acquired at 50$\%$ confluency is statistically significant when compared with the data at 60$\%$ confluency --- P$<$0.001. We note that without spectral decomposition, the total autofluorescence emission (Fig. \ref{fig:ResultsCombined}d) does not possess the same significant changes from 50$\%$ to 60$\%$ confluency. This transition in RR is therefore unique and can serve as a marker during cell manufacturing that can aid the operator in two ways: (i) allow objective prediction of a critical time for subculture, that is typically prescribed at 70$\%$ to 80$\%$ confluency; (ii) determine quality of cells based on growth rate (from the  time required to reach 60$\%$ confluency), as an early indicator for a go-no-go evaluation. 

Results show that RR is generally higher for lower confluencies ($\leq$50$\%$). We attribute this observation to the different type of metabolic requirements necessary within cells as they switch from individual survival to concerted proliferation\cite{VanderHeiden2009}. In the former, oxidative metabolism dominates as glucose is consumed for the generation of biomass and production of energy --- in the form of adenosine 5'-triphosphate (ATP). We expect this to happen for freshly seeded cells, as they adapt to the new environment and begin forming connections with the substrate and each other. Oxidative metabolism consumes NADH and generates FAD causing RR to increase and has been reported to be a hallmark of differentiated cells\cite{Quinn2012,Folmes2012,Croce2014}. Subsequently, as the cells enter a proliferative state they switch to anaerobic metabolism, where glucose is broken down for the creation of other precursors required in cell replication\cite{VanderHeiden2009}, instead of solely for the generation of energy. In contrast to oxidative metabolism, anaerobic metabolism increases the amount of NADH and decreases FAD, resulting in a lower RR. Lower RR has also been linked to cell proliferation\cite{Rice2010} and higher anabolic activities\cite{Quinn2012}. 

Corresponding changes in mitochondrial organisation were also noted from the autofluorescence images captured. Figure \ref{fig:ResultsCombined}f depicts the typical morphologies of cells and their mitochondria at different confluencies. We observed the transition of mitochrondrial organisation from being fragmented and distributed (for $\leq$30$\%$ and 40$\%$) to fragmented but concentrated around the nucleus (at 50$\%$) to interconnected and distributed (for $\geq$60$\%$). Fragmented mitochondria exist in cells entering division\cite{Friedman2014} and in cells that are less metabolically active\cite{Westermann2012}. These correspond to the higher RRs observed at $\leq$50$\%$ confluencies, which concurs with the reported inverse relationship between RR and metabolic activity\cite{Rice2010}. Conversely, mitochondria become interconnected in a fused and branched organisation when cells are metabolically active\cite{Westermann2012}. This has also been related to health and bioenergetic efficiency\cite{Picard2013} as well as used as an indicator of differentiation\cite{Quinn2013}.

\begin{table}[ht]
\caption{NADH and FAD concentration (per cell) computed from results of bio-chemical assays. ($^{\dagger}$Confluencies are reported in ranges due to variations across each 175cm$^2$ culture vessel.)} 
\label{tab:AssayResults}
\begin{center}       
\begin{tabular}{|c|c|c|c|}
\hline
\rule[-1ex]{0pt}{3.5ex}  \textbf{Confluency$^{\dagger}$ ($\%$)} 	& \textbf{[NADH]/cell ($\times$10$^{-5}$M)}	& \textbf{[FAD]/cell ($\times$10$^{-6}$M)}		& \textbf{Redox Ratio} \\
\hline\hline
\rule[-1ex]{0pt}{3.5ex}  40--50 							& 1.797$\pm$0.460						& 4.943$\pm$2.013						& 0.2157$\pm$0.0997 \\
\hline 
\rule[-1ex]{0pt}{3.5ex}  50--60 							& 7.084$\pm$0.750						& 1.941$\pm$0.632						& 0.0267$\pm$0.0091 \\
\hline 
\rule[-1ex]{0pt}{3.5ex}  80--90 							& 5.787$\pm$1.485						& 0.011$\pm$0.001						& 0.0002$\pm$0.0001 \\
\hline 
\rule[-1ex]{0pt}{3.5ex}  $\geq$90						& 8.261$\pm$1.976						& 0.138$\pm$0.111						& 0.0017$\pm$0.0014 \\
\hline 
\end{tabular}
\end{center}
\end{table} 

We further determined RR from bio-chemical assays, where the concentrations of NADH and FAD per cell were quantified bio-chemically at different confluencies as summarized in Table \ref{tab:AssayResults}. These data are also plot together with RR determined from autofluorescence spectroscopy in Fig. \ref{fig:Assays}a. It should be noted that confluencies were less specifically determined here because of variations across the large 175cm$^2$ area on which the cells were cultured --- in contrast to the $\sim$2cm$^{2}$ area in our silicone wells for microspectroscopy. We note from these results that RR is high for the confluency range of 40--50$\%$ and decreases at higher confluencies. This concurs well with the changes in RR observed via microscpectroscopy, where RR is significantly higher for confluencies of $\leq$50$\%$. We also note that RR is generally lower here as compared to those obtained via microspectroscopy, and have attributed this to the emission contributions of other flavin derivatives (flavin mononucleotides and riboflavins) which overlap with the emission band of FAD. Here, we wish to also highlight that each assay-based measurement destructively consumes at least 1$\times$10$^6$ cells and takes approximately 4hr to run. On the contrary, each set of microspectroscopy-based measurement can be completed non-destructively in under 15min --- inclusive of sample preparation, $\geq$10 spectral acquisitions and background measurements.

Following microspectroscopy, we performed a live-dead assay to ascertain that the duration of exposure to illumination did not compromise cell viability. The percentage of live cells was determined for both the illuminated cell samples and corresponding controls. Cells in the control came from the same batch of culture and were prepared and exposed to the same treatments with the exception of any illumination from the fluorescence microscope. Results from a 4-day long study are shown in Fig. \ref{fig:Assays}. Through a statistical analysis between the viabilities of the samples and their corresponding controls, we determined that there was no statistical significance between each pair. We thus inferred that the illumination did not affect cell viability. Nevertheless, we acknowledge the adverse effects that photochemical degradation (especially in flavins\cite{Rosner2016}) can have on cells. This can typically be mitigated through the use of pulsed laser sources or more sensitive detectors, both of which reduces the effective exposure of cells to illumination. On the other hand, we do note statistical significance between the viability of samples cultured for different durations. In particular, there was significant difference between the sample at 40$\%$ confluency (1 day in culture) and samples at 60$\%$ and 80$\%$ confluencies (2 and 3 days in culture respectively). We attribute this to cells being less susceptible to death at higher confluencies when they are acclimatized, healthier and more metabolically active --- discussed earlier to be a trait at higher confluencies and lower RR. 

\section{Conclusion}
\label{sect:conclusion}  
In this work, we used autofluorescence spectroscopy to determine the redox ratios (RR) in cells at different confluencies. Autofluorescence spectra were acquired from cells through an inverted fluoresence microscope with simple upgrades. Through spectral decomposition of these spectra we were able to determine relative compositions of NADH and FAD in cells and consequently the relative extents of reduction and oxidation respectively --- as represented by RR. Results showed significantly higher RR for confluencies of $\leq$50$\%$, which concurred well with RR determined through bio-chemical assays. We attributed this to cells at lower confluencies mainly undergoing oxidative metabolism for the generation of biomass and energy required in individual growth; while cells at higher confluencies primarily experience anaerobic metabolism for the conversion of glucose into other precursors required in proliferation. Through autofluorescence imaging, we further observed changes in mitochrondrial organisation that corresponded to the inverse relationship between metabolic activity and RR, where at high confluencies (lower RR) we noted highly interconnected mitochrondria that indicated high metabolic activity. Live-dead studies following our spectral acquisitions revealed that the exposure to illumination during our measurements had no significant effects on cell viability. In conclusion, we demonstrated a non-destructive method of determining cell confluency. The optical spectroscopy basis of the method allows it to be readily extended to meet the \textit{in situ} and real-time requirements of monitoring in cell therapy manufacturing. 

\appendix   

\subsection*{Disclosures}
The authors declare no conflict of interest.

\acknowledgments 
This work was supported by the Agency for Science Technology and Research (A$^*$STAR), Singapore. We thank our interns - Amanda Chia, Isaac Tan, Lee Pei Pei, Xie Yumin, Darren Chang and Lucas Foo - for their assistance in this work. 

%%%%% References %%%%%

%%%%% Biographies of authors %%%%%

\vspace{2ex}\noindent\textbf{Derrick Yong} received his BEng and PhD in Bioengineering from Nanyang Technological University, Singapore, in 2011 and 2016 respectively. He was awarded a graduate scholarship by the Agency for Science Technology and Research (A*STAR), Singapore, for his graduate research at the Singapore Institute of Manufacturing Technology, A*STAR. Since 2015, he has joined the same institute as a research scientist, studying bio-photonics and its applications in bio-manufacturing. His research interests include spectroscopy for biology and bio-integrated photonics.

\vspace{2ex}\noindent\textbf{Ahmad Amirul Abdul Rahim} received his BEng in Bioengineering from Nanyang Technological University, Singapore, in 2014. Currently, he is a research engineer at the Singapore Institute of Manufacturing Technology, developing devices that interface engineering and biology. His research interests include product and engineering design for biomedical products specifically in cell and tissue processing and manufacturing.

\vspace{2ex}\noindent\textbf{Jesslyn Ong} is a lab officer in the Bio-Manufacturing Programme, with concentration in cell culture and molecular biology, at the Singapore Institute of Manufacturing Technology, A*STAR. She graduated with a Diploma in Biotechnology and has experience in the pharmaceutical manufacturing industry. Her research interests include bio-chemical and structural biology, cell biology and infectious diseases.

\vspace{2ex}\noindent\textbf{May Win Naing} heads the Bio-Manufacturing Programme at the Singapore Institute of Manufacturing Technology, A*STAR. She received her BEng in Mechanical $\&$ Production Engineering and PhD in tissue engineering from Nanyang Technological University, Singapore. Prior to joining A*STAR in 2013, she has worked at the EPSRC Centre for Manufacturing of Regenerative Medicine at Loughborough University, UK, and has also taken on R$\&$D and Marketing roles in the medical device industry, specializing in spinal implants. Having worked in both academic and industry settings in Singapore and abroad, she is committed to the translation of technologies into the clinic and the market. Her research interest centers on scale-up manufacturing of biological products such as tissue scaffolds and cell therapies for applications in regenerative medicine, toxicity testing and consumer products.

%\vspace{1ex}
%\noindent Biographies and photographs of the other authors are not available.

\listoffigures
\listoftables

\end{spacing}
\end{document}